\documentclass[11pt]{article}
\usepackage{amssymb}
\usepackage{latexsym}
\begin{document}
\setlength{\baselineskip}{15pt}
\title{Generalized results on the role of new-time transformations \\
in finite-dimensional Poisson systems}
\author{ \mbox{} \\ Benito Hern\'{a}ndez-Bermejo $^1$}
\date{}

\maketitle
\begin{center}
{\em Departamento de F\'{\i}sica. Escuela Superior de Ciencias Experimentales y 
Tecnolog\'{\i}a. \\ 
Universidad Rey Juan Carlos. Calle Tulip\'{a}n S/N. 28933--M\'{o}stoles--Madrid. Spain.} 
\end{center}

\mbox{}

\mbox{}

\mbox{}

\begin{center} 
{\bf Abstract}
\end{center}
\noindent
The problem of characterizing all new-time transformations preserving the Poisson structure of a finite-dimensional Poisson system is completely solved in a constructive way. As a corollary, this leads to a broad generalization of previously known results. Examples are given. 

\mbox{}

\mbox{}

\mbox{}

\mbox{}



\mbox{}


\vfill

\noindent $^1$ Telephone: (+34) 91 488 73 91. Fax: (+34) 91 664 74 55. \newline 
\mbox{} \hspace{0.05cm} E-mail: benito.hernandez@urjc.es 

\pagebreak
\begin{flushleft}
{\bf 1. Introduction}
\end{flushleft}

Finite-dimensional Poisson systems have a significant presence in most fields of nonlinear physics, including domains such as mechanics, electromagnetism, optics, plasma physics, control theory, population dynamics, etc. (for instance, see \cite{olv1} for an overview and a historical discussion). Actually, recasting a given vector field as a Poisson system allows the use of a variety of techniques and specific methods adapted to that format, embracing stability analysis, perturbation methods, bifurcation analysis and characterization of chaotic behavior, efficient numerical integration, or integability properties and determination of invariants, just to mention a sample (e.g. see the discussions in \cite{bn4,bma1} for a brief account of these methods and application domains). Very diverse physical systems have been reported to be of the Poisson kind (a sample is given in \cite{gyn1}-\cite{dam1} and references therein) in spite that such identification often constitutes a nontrivial issue. 

When expressed in coordinates $x_1, \ldots , x_n$, a smooth dynamical system defined in a domain $\Omega \subset \mathbb{R}^n$ is said to be a finite-dimensional Poisson system if it can be written in the form
\begin{equation}
    \label{nd1nham}
    \dot{x} \equiv \frac{\mbox{d}x}{\mbox{d}t}= {\cal J}(x) \cdot \nabla H(x) 
\end{equation} 
where $x = (x_1, \ldots ,x_n)^T$, superscript $^T$ denotes the transpose of a matrix, function $H(x)$ is by construction a time-independent first integral (the Hamiltonian), and the $n \times n$ structure matrix ${\cal J}(x)$ is composed by the structure functions $J_{ij}(x)$ which must verify the Jacobi PDEs: 
\[
     \sum_{l=1}^n 
	\left( \begin{array}{c} J_{li}(x) \partial_l J_{jk}(x) + J_{lj}(x) \partial_l J_{ki}(x) + 
     J_{lk}(x) \partial_l J_{ij}(x) \end{array} \right) = 0 
	\:\; , \;\:\;\: i,j,k=1, \ldots ,n
\]
where $ \partial_l \equiv \partial / \partial x_l$. The structure functions must be also 
skew-symmetric: $J_{ij}(x) =  - J_{ji}(x)$ for all $i,j=1, \ldots ,n$. 

The relevance of Poisson dynamical systems relies on several reasons. One is that they provide a broad generalization of classical Hamiltonian systems, allowing not only for odd-dimensional vector fields, but also because Poisson structure matrices admit a great diversity of forms apart from the classical (constant) symplectic matrix. Actually, Poisson systems are a generalization of classical Hamiltonian systems on which a generalized Poisson bracket is defined, namely:
\begin{equation}
\label{nd1cp}
	\{ f(x),g(x) \} = \sum _{i,j=1}^n \partial_i f(x) J_{ij}(x) \partial_j g(x)
\end{equation}
for every pair of smooth functions $f(x)$ and $g(x)$. When Rank($\cal{J}$)$=n$ the Poisson system is termed symplectic. The possible rank degeneracy of the structure matrix ${\cal J}$ implies that a certain class of first integrals ($C(x)$ in what follows) termed Casimir invariants exist. There is no analog in the framework of classical Hamiltonian systems for such constants of motion, which are characterized by the property of having a null Poisson bracket ---in the sense of (\ref{nd1cp})--- with all smooth functions defined in $\Omega$. As it can be seen, this implies that Casimir invariants are the solution set of the system of coupled PDEs: ${\cal J} \cdot \nabla C =0$. The determination of Casimir invariants and their use in order to carry out a reduction (local, in principle) is the cornerstone of the (at least local) equivalence between Poisson systems and classical Hamiltonian systems, as stated by Darboux' theorem \cite{olv1}. This justifies that Poisson systems can be regarded, to a large extent, as a natural generalization of classical Hamiltonian systems. 

Among the different open issues in Poisson systems theory, a relevant topic arises in connection with the role of time in the preservation of a Poisson structure for the system. More precisely, a new-time transformation (or NTT in what follows) is defined as a reparametrization of the form
\begin{equation}
\label{ntt0}
	\mbox{\rm d}\tau = \frac{1}{\eta (x)}\mbox{\rm d}t
\end{equation}
where $t$ is the initial time variable, $\tau$ is the new time and $\eta (x) : \Omega \subset 
\mathbb{R}^n \rightarrow \mathbb{R} - \{ 0 \}$ is a smooth function in $\Omega$ which does not vanish in $\Omega$. Thus, if (\ref{nd1nham}) is an arbitrary Poisson system defined in $\Omega$, then the NTT (\ref{ntt0}) leads from 
(\ref{nd1nham}) to the system:
\begin{equation}
\label{ndposntt}
	\frac{\mbox{\rm d}x}{\mbox{\rm d} \tau} = \eta (x){\cal J}(x) \cdot \nabla H (x)
\end{equation}
NTTs play an interesting role in the framework of Poisson systems theory. For instance, they are required frequently in order to achieve reductions to the Darboux canonical form 
\cite{bn4,bma1,bs2},\cite{bn1}-\cite{bn5}. Additionally, NTTs can be found in the construction of Poisson structures in different physical problems \cite{mm1}-\cite{agfs}. Moreover, in the particular (but prominent) case of classical Hamiltonian systems, the use of NTTs is well-known for the analysis of integrability (specially in the context of the Painlev\'{e} property \cite{hgdr}-\cite{gor1}), in stability theory \cite{mha}, etc. Of course, time reparametrizations do not alter the form of trajectories in phase space, their only dynamical role amounts to modifying the ``speed'' at which every point moves on the system trajectories. However, now equations (\ref{ndposntt}) are not necessarily of Poisson type, namely NTTs do not preserve in general the Poisson format of the system. In a recent work \cite{bma1} this issue was analyzed from the following point of view: provided ${\cal J}$ is a structure matrix, it is intended to identify the cases in which $\eta {\cal J}$ remains a structure matrix. In \cite{bma1}, the following three main results were found: 

{\em Let ${\cal J}(x)$ be an $n \times n$ structure matrix of constant rank $r$ defined in a domain $\Omega \subset \mathbb{R}^n$, let $C(x)$ be a Casimir invariant of ${\cal J}(x)$ in $\Omega$, and let $\eta (x)$ be a smooth function in $\Omega$. Then: 
\begin{itemize}
\item Result 1: The product $\eta (x) {\cal J}(x)$ is a structure matrix in $\Omega$ for every smooth function $\eta (x)$ if and only if $r \leq 2$.
\item Result 2: $C(x) {\cal J}(x)$ is a structure matrix everywhere in $\Omega$.
\item Result 3: For $n \geq 4$, if ${\cal J}(x)$ is symplectic (namely $r=n$) then the product $\eta (x) {\cal J}(x)$ is a structure matrix in $\Omega$ if and only if $\eta (x)$ is a constant. 
\end{itemize}
}

Thus the results in \cite{bma1}, just summarized, consider the existence of a Poisson structure in the transformed system, but such Poisson structure (associated with the rescaled structure matrix $\eta {\cal J}$) is different from the initial one (corresponding to $\cal{J}$). In this sense, it is very natural to investigate the issue of the strict invariance of the Poisson structure after an NTT. Namely, we now aim at characterizing those NTTs that not only preserve the existence of a (possibly rescaled, and thus different) Poisson structure, but instead those that preserve the existence of the original Poisson structure. It is not difficult to realize that this complementary problem amounts to a rescaling of the Hamiltonian while leaving the Poisson structure (and thus the structure matrix) intact. Therefore, the purpose of this letter is to characterize the NTTs for which the reparametrized system (\ref{ndposntt}) can actually be written as: 
\begin{equation}
\label{recst}
	\frac{\mbox{\rm d}x}{\mbox{\rm d} \tau} = \eta (x) {\cal J}(x) \cdot \nabla H(x) = 
	{\cal J}(x) \cdot \nabla H^*(x) 
\end{equation}
for a new Hamiltonian $H^*(x)$. This problem is completely solved in what follows. Moreover, when combined with the complementary results developed in \cite{bma1}, already mentioned, an additional outcome will be a broad generalization of the conditions that an NTT must verify in order to preserve the existence of a (possibly rescaled) Poisson structure. 

The structure of the article is the following. In Section 2 the problem is solved for symplectic Poisson systems. The general solution of the problem is constructed in Section 3. Some examples are presented in Section 4. The letter concludes in Section 5 with some final remarks. 

\mbox{}

\begin{flushleft}
{\bf 2. General solution for symplectic Poisson systems}
\end{flushleft}

Both for the relevance of symplectic systems, as well as for the sake of clarity, it is very convenient to first consider the symplectic case independently. This provides a necessary starting point for the general treatment to be presented in Section 3. Recall that a Poisson system is termed symplectic if Rank($\cal{J}$)$=n$, which in turn implies that the system dimension $n$ is even. Thus, classical Hamiltonian systems are a particular (but important) case of symplectic vector fields. The fundamental result of this section is the following one: 

\mbox{}

\noindent{\bf Theorem 1.} 
{\em Let (\ref{nd1nham}) be an $n$-dimensional and symplectic Poisson system in a domain $\Omega \subset \mathbb{R}^n$. Then the NTT (\ref{ntt0}) preserves the Poisson structure of the system if and only if $\eta(x)$ and the Hamiltonian $H(x)$ are functionally dependent in $\Omega$. 
}

\mbox{}

\noindent{\bf Proof.} 
The proof is constructive. According to (\ref{recst}) we must have $\eta (x) {\cal J}(x) \cdot \nabla H(x) = {\cal J}(x) \cdot \nabla H^*(x)$ for some $H^*(x)$. Since the system is symplectic, ${\cal J}(x)$ is invertible and the problem amounts to $\eta (x) \nabla H(x) = \nabla H^*(x)$ for some new Hamiltonian $H^*(x)$. Now it is well-known (e.g. see \cite[Ch. 5]{frk1}) that $\eta (x) \nabla H(x)$ is a gradient if and only if $\partial_i( \eta (x) \partial_j H(x))=\partial_j( \eta (x)\partial_i H(x))$ for all $i,j=1, \ldots ,n$. After some algebra such conditions lead to: 
\[
	\frac{\partial _i \eta (x)}{\partial _j \eta (x)} = \frac{\partial _i H (x)}{\partial _j	H (x)} \;\: , \:\;\:\;\: i,j = 1, \ldots ,n
\]
These conditions imply that the Jacobian of $\eta (x)$ and $H(x)$ has rank 1 everywhere in $\Omega$, which in turn shows the functional dependence of $\eta (x)$ and $H(x)$. $\:\;\:\; \Box$

\mbox{}

Accordingly, if the symplectic Poisson structure is to be preserved by the NTT, there exists a 
one-variable smooth function $\phi (z)$ such that $\eta (x) = \phi(H(x))$. Let $\Phi (z)$ be a primitive of $\phi (z)$, then we conclude that the new Hamiltonian is $H^*(x) = \Phi (H(x))$, in such a way that: $\nabla H^*(x) = \nabla \Phi (H(x)) = \Phi ' (H(x)) \nabla H(x) = \phi (H(x)) \nabla H(x)$. This shows that the result is natural as a sufficient condition. The fact that Theorem 1 proves also that the condition is necessary is probably less evident. 

It is worth emphasizing that, in particular, Theorem 1 embraces classical Hamiltonian systems. In this context, two different aspects can be mentioned, namely the generic Poisson point of view and the purely Hamiltonian perspective: 

\begin{itemize}
\item From the Poisson point of view, the results in \cite{bma1} show that a Hamiltonian system can be submitted to an NTT in such a way that the outcome will be a Poisson (but generically not a Hamiltonian) system. In this sense, a one degree of freedom ($n=2$) Hamiltonian system can be reparametrized with every NTT, while for a two or more degrees of freedom ($n=4,6, \ldots $) Hamiltonian system, only constant NTTs were known. As indicated, in both cases the result is a Poisson but not a Hamiltonian system (the only exception being the trivial NTT for which $\eta (x)=1$). These are the conclusions that can be obtained by means of the results presented in \cite{bma1}. 
\item In the framework of Theorem 1 it is now possible to adopt a purely Hamiltonian perspective, and consider NTTs in which both the initial and the reparametrized systems are Hamiltonian. The outcome is then that a necessary and sufficient condition for this is $\eta (x) = \phi (H(x))$. To the author's knowledge, this result is known in the literature as a sufficient condition since \cite{cgl1}, and as a necessary and sufficient condition since \cite{igmg}, in both cases only for two-dimensional (one degree of freedom) Hamiltonian flows. 
\end{itemize}

We can now proceed to the analysis of the problem in its full generality. This is the aim of the next section. 

\mbox{}

\begin{flushleft}
{\bf 3. General solution of the problem}
\end{flushleft}

In the general case, the structure matrix needs not be symplectic. There is a simple heuristic argument that clearly points out that the result in Theorem 1 is not the general solution of the problem when Casimir invariants are present. For this, consider a Poisson system $\dot{x}= {\cal J}(x) \cdot \nabla H(x)$. If we rescale the Hamiltonian as $H^*(x) = C(x) H(x)$ with $C(x)$ being a Casimir invariant of the structure matrix ${\cal J}(x)$, then the new system remains as a Poisson one, namely $\dot{x}= {\cal J}(x) \cdot \nabla (C(x)H(x))$. However, this implies that: 
\[
\dot{x} = {\cal J}(x) \cdot \nabla (C(x)H(x)) = {\cal J}(x) \cdot ( C(x) \nabla H(x) + H(x) \nabla C(x)) = C(x) {\cal J}(x) \cdot \nabla H(x)
\]
And therefore it is clear that such rescaling of the Hamiltonian (which is thus equivalent to a rescaling of the structure matrix) must preserve the existence of a Poisson structure. This is another way of regarding the Result 2 from \cite{bma1} recalled in the Introduction. Moreover, this simple argument shows that in the nonsymplectic case, some NTTs are possible in which the structure matrix remains intact, and $\eta (x)$ is not only a function of the Hamiltonian. After these considerations, the following result should be natural: 

\mbox{}

\noindent{\bf Theorem 2.} 
{\em Let (\ref{nd1nham}) be an $n$-dimensional Poisson system with structure matrix of constant rank $r$ in a domain $\Omega \subset \mathbb{R}^n$. Let $C_1(x), \ldots , C_{n-r}(x)$ be a complete set of independent Casimir invariants of ${\cal J}(x)$ in $\Omega$. Consider an arbitrary implicit functional relationship of the form $F(H(x),H^*(x),C_1(x), \ldots , C_{n-r}(x))=0$ for all $x \in \Omega$, such that $F(z_1, \ldots ,z_{n-r+2})$ is smooth, and verifies that $\partial _{z_1} F \neq 0$ and $\partial _{z_2} F \neq 0$ everywhere. Then there exists in $\Omega$ a unique smooth Hamiltonian of the form $H^*(x) = \Phi (H(x),C_1(x), \ldots , C_{n-r}(x))$, as well as a smooth function 
\begin{equation}
\label{etath2}
\eta (x) = - \frac{ \partial _H F(H,H^*,C_1, \ldots , C_{n-r})}{\partial _{H^*} F(H,H^*,C_1, \ldots , C_{n-r})}
\end{equation}
defining an NTT of the form (\ref{ntt0}), such that 
\[
	\frac{\mbox{\rm d}x}{\mbox{\rm d} \tau} = \eta (x) {\cal J}(x) \cdot \nabla H(x) = 
	{\cal J}(x) \cdot \nabla H^*(x) 
\]
}


\noindent{\bf Proof.} 
If there exists a functional relationship of the form $F(H(x),H^*(x),C_1(x), \ldots , C_{n-r}(x))=0$ with $F(z_1, \ldots ,z_{n-r+2})$ smooth, and verifying $\partial _{z_2} F \neq 0$, then the implicit function theorem implies the existence of a unique and smooth function $H^*(x) = \Phi (H(x),C_1(x), \ldots , C_{n-r}(x))$. Moreover, after a differentiation of the implicit expression we find: 
\[
	\nabla F(H,H^*,C_1, \ldots , C_{n-r}) = 
	(\partial _H F) \nabla H + (\partial _{H^*} F) \nabla H^* + 
	\sum_{i=1}^{n-r} (\partial _{C_i}F) \nabla C_i = 0
\]

Multiplying by the structure matrix, and taking into account that ${\cal J} \cdot \nabla C_i =0$ for all the Casimir invariants, we arrive at:
\[
	{\cal J} \cdot \nabla H^* 
	 = - \left( \frac{\partial _H F}{\partial _{H^*} F} \right) {\cal J} \cdot \nabla H 
\]
Now this expression corresponds to an NTT applied to the original Poisson system provided $\eta (x) \neq 0$ everywhere, which is guaranteed by the additional condition $\partial _{z_1} F \neq 0$. $\:\;\:\; \Box$

\mbox{}

In spite of its generality, this result does only prove the existence of a unique rescaled Hamiltonian, but it is not constructive. From an operational point of view, it is desirable to have an explicit, and thus constructive result, rather than an implicit one. This is done in the next theorem, which is the most important of this section: 

\mbox{}

\noindent{\bf Theorem 3.} 
{\em Let (\ref{nd1nham}) be an $n$-dimensional Poisson system with structure matrix of constant rank $r$ in a domain $\Omega \subset \mathbb{R}^n$. Let $C_1(x), \ldots , C_{n-r}(x)$ be a complete set of independent Casimir invariants of ${\cal J}(x)$ in $\Omega$. Then the NTT (\ref{ntt0}) preserves the Poisson structure of the system if and only if $\eta(x)$ is functionally dependent on the Hamiltonian $H(x)$ and the Casimir invariants in $\Omega$, namely if and only if there exists in $\Omega$ a smooth dependence of the form $\eta(x) = \phi (H(x), C_1(x), \ldots , C_{n-r}(x))$. If this is the case, let $\Phi (z_1, \ldots , z_{n-r+1})$ be a function such that $\partial_{z_1} \Phi (z_1, \ldots , z_{n-r+1})= \phi (z_1, \ldots , z_{n-r+1})$. Then, it is 
\[
	\frac{\mbox{\rm d}x}{\mbox{\rm d} \tau} = \eta (x) {\cal J}(x) \cdot \nabla H(x) = 
	{\cal J}(x) \cdot \nabla H^*(x) 
\]
with $H^*(x) = \Phi (H(x), C_1(x), \ldots , C_{n-r}(x))$. 
}

\mbox{}

\noindent{\bf Proof.} 
In one sense, consider the new Hamiltonian $H^*(x) = \Phi (H(x), C_1(x), \ldots , C_{n-r}(x))$. By differentiation we find: 
\[
	\nabla H^* = ( \partial_H \Phi ) \nabla H + \sum_{i=1}^{n-r} ( \partial _{C_i} \Phi )	\nabla C_i
\]
Since it is ${\cal J} \cdot \nabla C_i =0$ for all the Casimir invariants, we obtain
\[
	{\cal J} \cdot \nabla H^* = (\partial_H \Phi ) {\cal J} \cdot \nabla H 
\]
and we thus have $\eta(x) = \phi (H(x), C_1(x), \ldots , C_{n-r}(x)) = 
\partial_H \Phi (H(x), C_1(x), \ldots , C_{n-r}(x))$.

In the opposite sense, let us prove that these are the most general possible forms for $H^*(x)$ and $\eta (x)$. For this, we shall investigate the most general possible form which the rescaled Hamiltonian $H^*(x)$ could have in principle. In the case of the most general dependence, it could be a function of $H(x)$ and the Casimir invariants $C_i(x)$, $i=1, \ldots ,n-r$, but also of the coordinates $x$ and of a set of additional functions $K_1(x), \ldots , K_m(x)$ which are functionally independent on $H(x)$ and the Casimir functions. Namely, the most general functional dependence possible in principle for the rescaled Hamiltonian is the following one: 
\begin{equation}
\label{gth3}
	G(x) = M (x_1, \ldots, x_n,H(x), C_1(x), \ldots , C_{n-r}(x), K_1(x), \ldots 	,K_m(x))
\end{equation}
In this equation, the function candidate to play the role of the rescaled Hamiltonian is termed $G(x)$ instead of $H^*(x)$ for reasons that will become apparent in brief. If we differentiate this expression, we readily find: 
\[
	\nabla G(x) = \nabla _x M + (\partial _H M) \nabla H + 
	\sum _{j=1}^{n-r} ( \partial _{C_j} M) \nabla C_j + 
	\sum _{l=1}^m ( \partial _{K_l} M) \nabla K_l 
\]
where $\nabla _x M$ denotes the column vector of entries $(\nabla _x M )_i = \partial _{x_i} M$ for all $i=1, \ldots, n$. Once we multiply both sides of this identity by the structure matrix, the outcome must be an expression of the form $\eta {\cal J} \cdot \nabla H = {\cal J} \cdot \nabla H^*$. Since the terms associated with the Casimir invariants cancel out, this is possible if and only if the expression
\[
	\nabla G(x) - \nabla _x M - \sum _{l=1}^m ( \partial _{K_l} M) 	\nabla K_l 
\]
is a gradient. Since $\nabla G(x)$ is already a gradient, this means that 
\[
	\nabla _x M + \sum _{l=1}^m ( \partial _{K_l} M) \nabla K_l 
\]
is also a gradient. In addition, the existence of the previous gradients implies that
\[
	(\partial _H M) \nabla H + \sum _{j=1}^{n-r} ( \partial _{C_j} M) \nabla C_j 
\]
is another gradient. Obviously, these facts imply that the most general functional form of function $M$ is actually the following one: 
\[
	G(x) = P (x_1, \ldots, x_n, K_1(x), \ldots ,K_m(x)) + \Phi (H(x), C_1(x), \ldots , 
	C_{n-r}(x))
\]
If we differentiate this identity, we arrive at: 
\[
	\nabla (G-P) = \nabla \Phi (H, C_1, \ldots , C_{n-r}) = 
	( \partial _H \Phi ) \nabla H + \sum_{j=1}^{n-r} ( \partial _{C_j} \Phi ) \nabla C_j
\]
Now if we multiply both sides by the structure matrix, the outcome is:
\[
	{\cal J} \cdot \nabla (G-P) = {\cal J} \cdot \nabla \Phi = 
	( \partial _H \Phi ) {\cal J} \cdot \nabla H 
\]
This means that, in fact, the actual rescaled Hamiltonian is not $G$, but $(G-P)= \Phi$. Notice then that $\Phi $ only depends functionally on $H(x)$ and the Casimir functions. In other words, we can write $H^*(x) = \Phi (H(x), C_1(x), \ldots , C_{n-r}(x))$. Thus, the assumed dependence on $x$ and on the functions $K_i$ vanishes, and the most general dependence of $H^*(x)$ is only on $H(x)$ and the Casimir invariants, as stated. Note also that it is $\eta = ( \partial _H \Phi )$. $\:\;\:\; \Box$

\mbox{}

Some remarks are convenient at this stage. In first place, it is interesting to note that the explicit result developed in Theorem 3 includes as a particular case the symplectic situation, as expected: the reason is that in the symplectic case there are no Casimir invariants, and then the most general rescaled Hamiltonian $H^*(x) = \Phi (H(x), C_1(x), \ldots , C_{n-r}(x))$ reduces to $H^*(x) = \Phi (H(x))$. Similarly, we obtain $\eta (x)= \partial _H \Phi (H(x),C_1(x), \ldots ,C_{n-r}(x))$ which consistently becomes $\eta (x)= \Phi '(H(x))$ in the symplectic case, as anticipated in Theorem 1. 

An additional remark involves the relationship between Theorems 2 and 3. Note that the explicit dependence can be regarded from the implicit perspective if we note that there is a functional dependence of the form $F(H,H^*,C_1, \ldots ,C_{n-r})=H^*- \Phi(H,C_1, \ldots ,C_{n-r})=0$. In such case, by applying Theorem 2 we find from (\ref{etath2}) that 
\[
\eta (x) = - \frac{ \partial _H F(H,H^*,C_1, \ldots , C_{n-r})}{\partial _{H^*} F(H,H^*,C_1, \ldots , C_{n-r})} = \partial _H \Phi(H,C_1, \ldots ,C_{n-r}) 
\]
which is consistently the outcome of Theorem 3. 

Clearly, the results just developed complement those from \cite{bma1} which were summarized in the Introduction. Together, they provide an overall generalization which can be presented at this stage. 

\mbox{}

\noindent{\bf Corollary 1.} 
{\em Let (\ref{nd1nham}) be an $n$-dimensional Poisson system with structure matrix of constant rank $r$ in a domain $\Omega \subset \mathbb{R}^n$. Let $C_1(x), \ldots , C_{n-r}(x)$ be a complete set of independent Casimir invariants of ${\cal J}(x)$ in $\Omega$. After the NTT (\ref{ntt0}) the outcome is another Poisson system of the form
\[
	\frac{\mbox{\rm d}x}{\mbox{\rm d} \tau} = \eta (x) {\cal J}(x) \cdot \nabla H(x) = 
	{\cal J}^*(x) \cdot \nabla H^*(x) 
\]
where ${\cal J}^*(x)$ is a new structure matrix and $H^*(x)$ is a new Hamiltonian, if one of the following cases hold: 
\begin{itemize}
\item If $r=2$, then ${\cal J}^*(x) = \eta _0(x) {\cal J}(x)$ for an arbitrary smooth nonvanishing function $\eta _0(x)$ and $H^*(x)= \Phi (H(x),C_1(x), \ldots ,C_{n-r}(x))$ for every smooth function $\Phi (z_1, \ldots, z_{n-r+1})$ such that $\partial _{z_1} \Phi \neq 0$ everywhere. In this case, it is $\eta (x) = \eta _0(x) \partial _H \Phi (H(x),C_1(x), \ldots ,C_{n-r}(x))$. 
\item If $r \geq 4$, then ${\cal J}^*(x) = C(x) {\cal J}(x)$ for an arbitrary nonvanishing Casimir invariant $C(x)$ and $H^*(x)= \Phi (H(x),C_1(x), \ldots ,C_{n-r}(x))$ for every smooth function $\Phi (z_1, \ldots, z_{n-r+1})$ such that $\partial _{z_1} \Phi \neq 0$ everywhere. We thus have $\eta (x) = C(x) \partial _H \Phi (H(x),C_1(x), \ldots ,C_{n-r}(x))$. 
\item If $r \geq 4$ and ${\cal J}(x)$ is symplectic, then ${\cal J}^*(x)=c {\cal J}(x)$, with $c$ a nonzero constant, and $H^*(x)= \Phi (H(x))$ for every smooth function $\Phi (z)$ such that $\Phi '(z)\neq 0$ everywhere. In this case, it is $\eta (x) = c \Phi '(H(x))$. 
\end{itemize}
}

\mbox{}

This completes the presentation of results. In the next section, some examples are given. 

\mbox{}

\begin{flushleft}
{\bf 4. Examples}
\end{flushleft}

\noindent {\bf Example 1.} {\em Classical Hamiltonian systems: the harmonic oscillator.}

\mbox{}

As indicated in Section 2, in the case of Hamiltonian systems the fact of performing an NTT preserving the existence of a (possibly different) Poisson structure, often does not preserve the Hamiltonian form of the equations of motion. According to Theorem 1, the outcome is also Hamiltonian if and only if $\eta (x) = \phi (H(x))$. It is illustrative to see how this result arises for a specific system. For this, the harmonic oscillator is chosen, namely $H(x)=(x_1^2+x_2^2)/2$. Accordingly, we have: 
\[
	\left( \begin{array}{c} \dot{x}_1 \\ \dot{x}_2 \end{array} \right) = 
	{\cal J} \cdot \nabla H =
	\left( \begin{array}{cc} 0 & 1 \\ -1 & 0 \end{array} \right) \cdot 
	\left( \begin{array}{c} x_1 \\ x_2 \end{array} \right) = 
	\left( \begin{array}{c} x_2 \\ -x_1 \end{array} \right) 
\]
If an NTT of the form (\ref{ntt0}) leads to a Hamiltonian system, this means that 
\[
	\left( \begin{array}{c} \eta (x_1,x_2) x_1 \\ \eta (x_1,x_2) x_2 \end{array} \right) = 
	\nabla H^*(x_1,x_2) =  
	\left( \begin{array}{c} \partial_1 H^*(x_1,x_2) \\ \partial_2 H^*(x_1,x_2) 
	\end{array} \right) 
\]
The left-hand side of this expression is a gradient if and only if $\partial_2(\eta (x_1,x_2) x_1)= \partial_1(\eta (x_1,x_2) x_2 )$. This expression is equivalent to the following PDE: $x_2 \partial_1 \eta (x_1,x_2)- x_1 \partial _2 \eta (x_1,x_2)=0$. The general solution of this PDE consists of all the functions $\eta (x_1,x_2)$ associated with NTTs that preserve the Hamiltonian character of the equations of motion. Making use of the characteristics method we have: 
\[
	\frac{\mbox{d}x_1}{x_2}= \frac{\mbox{d}x_2}{-x_1} \;\: , \;\:\;\: \mbox{d} \eta =0
\]
We thus find two quadratures, namely: $c_1= x_1^2 +x_2^2$, and $c_2=\eta$. This implies that the general solution is of the form $\eta (x_1,x_2)= f(x_1^2+x_2^2)$. Taking into account the functional form of $H(x_1,x_2)$, this is equivalent to $\eta (x_1,x_2)= \phi(H(x_1,x_2))$. We thus see how this result can of course be developed from direct calculations, in spite that it arises in a simpler and more general way in the context of Theorem 1. 

\mbox{}

\noindent {\bf Example 2.} {\em Three-dimensional Poisson systems.}

\mbox{}

It is interesting to display an example showing when, and how, Theorem 3 works. Two cases will be illustrated: when the theorem is applicable and when it is not. For this, we shall consider a general 3-d Poisson system $\dot{x}= {\cal J}(x) \cdot \nabla H(x)$, with $x = (x_1,x_2,x_3)^T$, defined in a domain in which Rank(${\cal J}$)$=2$. This is the generic situation of interest in the 3-d case. In fact, the number of 3-d Poisson systems of rank two reported and analyzed in the literature is very significant: there are instances corresponding to the Euler top, the Kermack-McKendrick model for epidemics, several integrable cases of the Lorenz system, the Lotka-Volterra equations from population dynamics, the May-Leonard model, different families of Quasi-Polynomial systems, the Halphen equations, the circle maps system, the Maxwell-Bloch equations, several formulations of the Rabinovich system, the RTW interaction equations, etc. For instance, see \cite{bn4,gyn1,bs2,tur} and references therein for a sample. In addition many Poisson systems defined for a general dimension $n$ are meaningful, in particular, in the case $n=3$. The developments presented in this example are valid for all these 3-d systems and can be applied directly to any of them. 

Since it is Rank(${\cal J}$)$=2$ everywhere, there is one independent Casimir invariant $C(x)$. Additionally, we introduce a generic auxiliary smooth function $K(x)$ independent on $H(x)$ and $C(x)$. No specific functional form is assigned to $K(x)$ for the sake of generality. 

Accordingly, as a first possibility we focus on an instance complying to all the conditions established in the proof of Theorem 3. For this, the following dependence is considered for function $G(x)$ in (\ref{gth3}):
\[
	G(x) = M (x,H(x),C(x),K(x)) = x_1^2x_2x_3K + HC^2 
\]
Then, it is $P(x,K)=x_1^2x_2x_3K$ and $\Phi(H,C)=HC^2$. Now it is clear that 
$\nabla G- \nabla P$ is a gradient, actually it is $\nabla (G-P)= \nabla \Phi (H,C) = C^2 \nabla H + 2HC \nabla C$. Thus the problem follows the lines given in the proof of Theorem 3, and we have $H^*(x)= \Phi (H(x),C(x))= H(x)C^2(x)$. Then, we find ${\cal J} \cdot \nabla H^* = {\cal J} \cdot (C^2 \nabla H + 2HC \nabla C)=C^2 {\cal J} \cdot \nabla H$. Thus the NTT associated with this specific form of $G(x)$ is the one defined in terms of function $\eta (x) = C^2(x)$.

As the second possibility, we choose an instance based on the same dynamical system, on the same Poisson structure, and on the same functions $C(x)$, $H(x)$ and $K(x)$, but this time defined in such a way that the requirements developed in the proof of Theorem 3 are not verified. For this, we now define an alternative function $\tilde{G}(x)$:  
\[
	\tilde{G}(x) = \tilde{M} (x,H(x),C(x),K(x)) = x_1^2x_2x_3K + x_1HC^2 
\]
We see that $\tilde{G}(x)$ is not of the form $P(x,K)+\Phi(H,C)$. Anyway, let us try to reproduce the same steps in order to arrive to an NTT preserving the Poisson structure. For convenience, let us write $\tilde{G}(x)=P(x,K)+ x_1 \Phi(H,C)$. By differentiation, we find: 
\[
	\nabla (\tilde{G}-P) - HC^2 \vec{e}_1 =
	x_1 (C^2 \nabla H + 2HC \nabla C) 
\]
with $\vec{e}_1 = ( 1 , 0 , 0 )^T$. From the point of view of characterizing all possible NTTs, we can write the most general form of this expression by multiplying both sides by a smooth nonvanishing function $\gamma (x)$:
\[
	\gamma \nabla (\tilde{G}-P) - \gamma HC^2 \vec{e}_1 =
	x_1 \gamma (C^2 \nabla H + 2HC \nabla C) 
\]
Thus, after multiplying by the structure matrix we obtain: 
\[
	{\cal J} \cdot \left[ \gamma \nabla (\tilde{G}-P) - \gamma HC^2 \vec{e}_1 \right] =
	(x_1 \gamma C^2) {\cal J} \cdot \nabla H 
\]
Therefore, this process seems to lead to an NTT defined in terms of function $\eta (x)= x_1 \gamma (x) C^2(x)$ and applied to the initial Poisson system. Recall that according to 
(\ref{recst}) the structure matrix ${\cal J}(x)$ must remain the same after the NTT. However, if this is the case then there must be a function $H^*(x)$ such that $\nabla H^*(x) = \gamma \nabla (\tilde{G}-P) - \gamma HC^2 \vec{e}_1 = x_1 \gamma \nabla (HC^2)$. But this is possible if and only if $x_1 \gamma = \xi (HC^2)$ for some smooth and nonvanishing one-variable function $\xi (z)$. In such case, we have $\eta (x)= x_1 \gamma (x) C^2(x) = \xi (HC^2) C^2$, which does correspond to the kind of NTTs that preserve the Poisson structure according to Theorem 3. For the rest of choices of function $\gamma (x)$ the outcome is that the expression $x_1 \gamma \nabla (HC^2)$ is not a gradient and therefore $H^*(x)$ does not exist. 

\mbox{}

\begin{flushleft}
{\bf 5. Final remarks}
\end{flushleft}

As mentioned in the Introduction, NTTs constitute a valuable tool for the analysis of Poisson systems in different domains. In this sense, the study of NTTs preserving the Poisson format becomes a logical issue that provides results of interest, as well as an increasingly rich perspective of finite-dimensional Poisson structures. In this letter NTTs that strictly preserve the Poisson structure have been characterized constructively. This problem is relevant not only from a fundamental point of view, but also because it is natural in different applied contexts. Additionally, the results reported are not limited in dimension or in rank of the Poisson structure, thus being remarkably general in scope. Moreover, the characterizations obtained have been unified with those previously developed in \cite{bma1}, thus leading to a broader perspective. Of course, the role of NTTs in the Poisson framework is not completely understood at present, and therefore this issue should deserve additional attention in the future.

\pagebreak


\begin{thebibliography}{99}
   \bibitem{olv1} P.J. Olver, Applications of Lie Groups to Differential 
      Equations, 2nd Ed. (Springer-Verlag, New York, 1993).
   \bibitem{bn4} B. Hern\'{a}ndez-Bermejo, Phys. Lett. A 372 (2008) 1009.
   \bibitem{bma1} B. Hern\'{a}ndez-Bermejo, J. Math. Anal. Appl. 344 (2008) 655. 
   \bibitem{gyn1} H. G\={u}mral, Y. Nutku, J. Math. Phys. 34 (1993) 5691.
   \bibitem{byv2} B. Hern\'{a}ndez-Bermejo, V. Fair\'{e}n, J. Math. Phys. 39 (1998) 6162. 
   \bibitem{bhqpa} G.B. Byrnes, F.A. Haggar, G.R.W. Quispel, Physica A 272 (1999) 99. 
   \bibitem{bs2} B. Hern\'{a}ndez-Bermejo, J. Math. Phys. 42 (2001) 4984. 
   \bibitem{tur} A. Ay, M. G\"{u}rses, K. Zheltukhin, J. Math. Phys. 44 (2003) 5688.
   \bibitem{dam1} P.A. Damianou, S.P. Kouzaris, J. Phys. A: Math. Gen. 36 (2003) 1385. 
   \bibitem{bn1} B. Hern\'{a}ndez-Bermejo, J. Math. Phys. 47 (2006) 022901 1. 
   \bibitem{bn2} B. Hern\'{a}ndez-Bermejo, Phys. Lett. A 355 (2006) 98.
   \bibitem{bn5} B. Hern\'{a}ndez-Bermejo, Appl. Math. Lett. 22 (2009) 187.
   \bibitem{mm1} A.V. Borisov, I.S. Mamaev, Reg. \& Chaot. Dyn. 7 (2002) 177.
   \bibitem{mm2} A.V. Borisov, I.S. Mamaev, A.A. Kilin, Reg. \& Chaot. Dyn. 7 (2002) 201.
   \bibitem{ar1} A. Ramos, J. Phys. A: Math. Gen. 37 (2004) 4821.
   \bibitem{fgs1} F. Fass\`{o}, A. Giacobbe, N. Sansonetto, Reg. \& Chaot. Dyn. 10 (2005) 
	267.
   \bibitem{agfs} E. Abado\u{g}lu, H. G\={u}mral, Physica D 238 (2009) 526. 
   \bibitem{hgdr} J. Hietarinta, B. Grammaticos, B. Dorizzi, A. Ramani, Phys. Rev. Lett. 53 
	(1984) 1707.
   \bibitem{rgb} A. Ramani, B. Grammaticos, T. Bountis, Phys. Rep. 180 (1989) 159.
   \bibitem{gor1} A. Goriely, Integrability and Nonintegrability of Dynamical Systems 
	(World Scientific, Singapore, 2001).
   \bibitem{mha} K.R. Meyer, G.R. Hall, Introduction to Hamiltonian Dynamical Systems and the 	N-Body Problem (Springer-Verlag, New York, 1992).
   \bibitem{frk1} T. Frankel, The Geometry of Physics, 2nd Ed. (Cambridge University Press, 
	Cambridge UK, 2004).
   \bibitem{cgl1} M. Castagnino, M. Gadella, L. P. Lara, Chaos, Solitons \& Fractals 30 (2006) 
	542. 
   \bibitem{igmg} I.A. Garc\'{\i}a, M. Grau, A survey on the inverse integrating factor, 
	arXiv:0903.0941v1 [math.DS] (2009). 
\end{thebibliography}
\end{document}